# Evidence for spin droplets (ferrons) formation in the heavy fermion metal CeB$_6$ with dynamic charge stripes.

A.N. Azarevich[1], O.N. Khrykina[1,2], N.B. Bolotina[2], V.G. Gridchina[1,2], A.V. Bogach[1], S.V. Demishev[3,1], V.N. Krasnorussky[3,1], S. Yu. Gavrilkin[4], A.Yu. Tsvetkov[4], N.Yu. Shitsevalova[5], V.V. Voronov[1], K.I. Kugel[6], A.L. Rakhmanov[6], S. Gabáni[7], K. Flachbart[7], N.E. Sluchanko[1]

[1]*Prokhorov General Physics Institute of the Russian Academy of Sciences, Vavilova 38, 119991, Moscow, Russia*

[2]*National Research Center Kurchatov Institute, Moscow, 123182 Russia*

[3]*Vereshchagin Institute for High Pressure Physics of the Russian Academy of Sciences, 142190, Troitsk, Moscow, Russia*

[4]*Lebedev Physical Institute, Russian Academy of Sciences, Moscow, Leninsky av. 53, 119991 Russia*

[5]*Frantsevich Institute for Problems of Materials Science, National Academy of Sciences of Ukraine, 03680 Kyiv, Ukraine*

[6]*Institute for Theoretical and Applied Electrodynamics of Russian Academy of Sciences, Izhorskaya str. 13, 125412 Moscow, Russia*

[7]*Institute of Experimental Physics of the Slovak Academy of Sciences, Watsonova 47, SK−04001 Košice, Slovakia*

**Abstract.**

The presented studies of resistivity ($\rho$), thermal conductivity ($\kappa$) and specific heat ($C$) at low temperature 1.8−7 K in magnetic field up to 90 kOe made it possible to detect for the first time the exponential field dependences $\rho(H)$, $\kappa^{-1}(H)$, $C(H) \sim \exp(-\mu_{\text{eff}}H/k_BT)$ of the charge transport and thermal characteristics in the so−called antiferroquadrupole (AFQ) phase of the archetypal heavy-fermion CeB$_6$ hexaboride. From magnetoresistance measurements it is shown that in the AFQ state the effective magnetic moment varies in the range $\mu_{\text{eff}}(T) = 1.4-1.9\mu_B$, and its value is very close to $\mu_{\text{eff}(\tau)}(T) \approx 2\mu_B$, derived from the field dependence of the relaxation time $\tau(H)$ observed in the heat capacity and thermal conductivity experiments. The phenomenological model proposed here allowed us to attribute the magnetic moments to spin droplets (ferrons), that appear in the bulk AFQ phase of CeB$_6$ crystals. The relevant electron phase separation at the nanoscale, manifested by dynamic charge stripes, that leads to the formation of ferrons, was revealed from the analysis of low−temperature X-ray diffraction experiments using the maximum entropy method. We argue that the Jahn−Teller collective mode of B$_6$ clusters is responsible for the charge stripe formation, which subsequently induces transverse quasi-local vibrations of Ce ions in the form of pairs and triples. These lead to $4f-5d$ spin fluctuations providing spin-polarons (ferrons) in the CeB$_6$ matrix.

## 1. Introduction

It has long been believed that rare earth (RE) heavy fermion (HF) compounds are homogeneous materials, where the single−ion Kondo effect is responsible for the reduction of localized magnetic moments of Ce, Sm, Eu, Tm, Yb and U ions, which leads to formation of *metallic* strongly correlated electron systems (SCES) with unusual magnetic and non−magnetic ground states, and to metal−insulator transition in the so-called *Kondo insulators* (see, for example, [1] and references therein). The quenched disorder in the crystal structure of SCES [1−4] is considered usually either in the light of Kondo-disorder model [5−6] or magnetic Griffiths phases with short−range order [2−3,7−8], which appear inside the paramagnetic state above the magnetic phase transition. On the contrary, it is well-established nowadays that the SCES in the families of high-$T_c$ cuprates (HTSC) and iron-based pnictides, and also in manganites, cobaltites, etc. are mostly *inhomogeneous materials* with electronic phase separation, which coexists with various types of disorder [9, 10, 11, 12]. Indeed, a numerous fundamental studies of manganites [13–16, 17], HTSC cuprates [18–20, 21], iron-based superconductors [22−27], chalcogenides [28], etc. made it possible to discover a diversity of physical phenomena universal to these SCES, which demonstrate static and dynamic charge and spin stripes, charge and spin density waves (CDW and SDW), nematic phases, and structural inhomogeneities [9, 29−35].

During last two decades the similarity in the magnetic phase diagrams of HTSC and Ce−based HF SCES was discussed in numerous studies (see e.g. [36−37] for $CeCoIn_5$ HF superconductor) emphasizing both common features and relative inhomogeneous phases associated with the interplay between superconductivity and various type of AFM order. The spin droplets introduced by non-magnetic dopants in quantum critical superconductors $CeCo(In_{1−x}Cd_x)_5$ were proposed in [37] being consonant with similar effects found earlier, e.g. in manganites [29]. Very recently dynamic charge stripes and sub-structural CDW have been found [38−39] in $CeB_6$ HF metal with an unusual antiferromagnetic ground state, and singularities of the electron density (ED) distribution were detected directly from precise X-ray diffraction (XRD) measurements by maximal entropy method (MEM). Taking into account that, from the strong−coupling perspective, stripes are a real-space pattern of microphase separation [40], the nanoscale visualization of the filamentary structure could be realized with the help of scanning tunneling microscopy (STM) and scanning tunneling spectroscopy (STS). However, STS and STM are essentially static and surface sensitive techniques and consequently dynamic stripes can only be detected if they are pinned by impurities. Indirect methods, like anisotropic charge transport measurements on single domain crystals [41], and combination of precise XRD measurements with dynamic conductivity studies (see e.g. [42−44] for $Tm_{1−x}Yb_xB_{12}$ Kondo insulators) are considered to be the most effective techniques for detecting the symmetry reduction below the transition to the fluctuating stripe phase.

Until recently it has been believed that $CeB_6$ is the archetypal example of magnetic Kondo lattice with a small Kondo temperature $T_K \sim 1$ K, which is comparable to the temperatures of two magnetic phase transitions: (*i*) at $T_Q \approx 3.2$ K to orbital ordering in the antiferroquadrupolar (AFQ, phase II in Fig.1a) 'magnetically hidden order' phase and (*ii*) at $T_N \approx 2.3$ K into an unusual Néel antiferromagnetic (AFM, phase III in Fig. 1a) ground state (see [45] for review). In [46], the electron nematic effect was discovered in the II (AFQ) phase of $CeB_6$ (Fig.1a) suggesting symmetry breaking in this HF compound having initially a simple cubic crystal structure. It is argued in [38], that instead of the single-ion behavior the regime of Griffiths phase is valid in $CeB_6$ in a very wide temperature range 5−800 K, where magnetization follows an exponential dependence $M \sim H(T − T_Q)^{−0.8}$ for different directions of the applied magnetic field (Fig.1b). It is noted [38] that in $CeB_6$ is the Griffiths phase regime observed in an extra wide temperature range corresponding to the absolute record value of the Griffiths temperature $T_G > 800$ K for the magnetic clusters in disordered systems. In addition, a power-law behavior of the magnetic contribution to resistivity $\rho_m(T) \sim T^{-0.4}$ was deduced in [38] (see also Fig.1b) in the temperature range 8–90 K far above the Kondo temperature $T_K \sim 1$ K and well below the crystal electric field splitting $\Delta_{CEF} \approx 530$ K of the ground state $^2F_{5/2}$ multiplet of cerium, which points in favor of the regime of weak localization of charge carriers in the nanoscale clusters of Ce ions in this SCES. The above-mentioned anomalies are accompanied by intense ferromagnetic (FM) fluctuations. Indeed, a low-energy collective FM mode in the magnetic excitation spectrum of $CeB_6$ was found in inelastic neutron-scattering experiments [47]. The results of electron spin resonance (ESR) [48,49], muon spin relaxation ($\mu$SR) spectroscopy [50,51], and Hall effect [52] studies of $CeB_6$ confirmed the appearance of ferromagnetic nanosize domains embedded in the $CeB_6$ matrix, which places $CeB_6$ much closer to FM instability than previously thought. It is worth noting, that the classical interpretation of physical properties of $CeB_6$ is based almost solely on the consideration of $Ce^{3+}$ localized magnetic moments with a certain orbital degree of freedom [53−55]. Thus, the itinerant contribution is often considered via the Kondo-type screening.

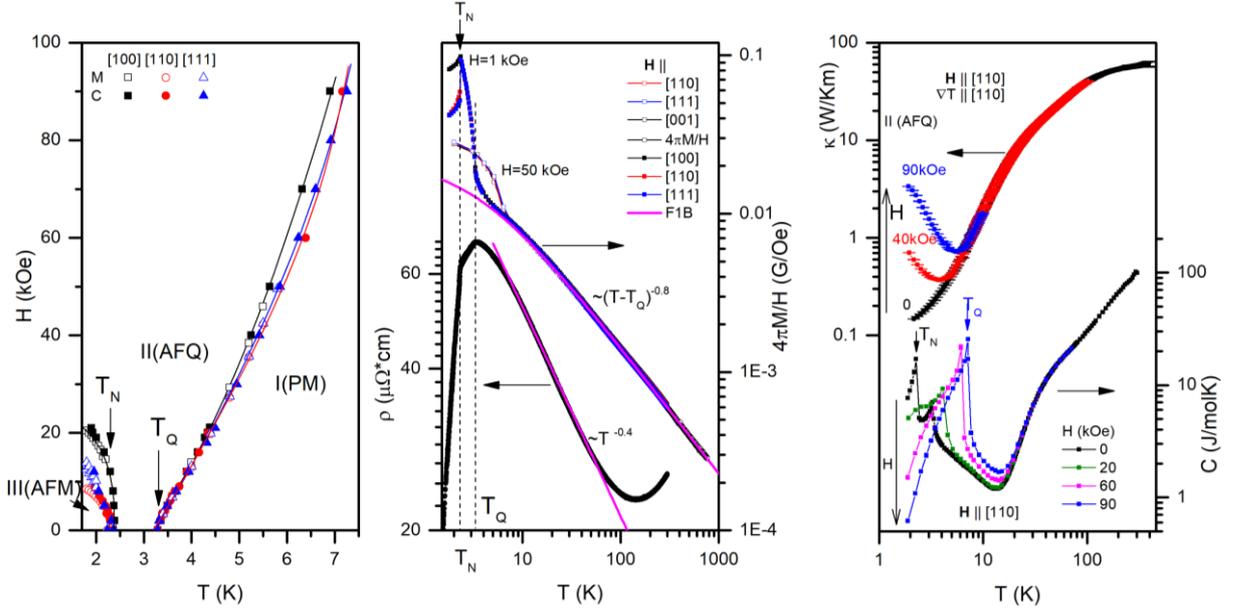

**Fig. 1.** (Color online) (a) Magnetic phase diagram of $CeB_6$. II (AFQ), III (AFM) and I (PM) denote the antiferroquadrupolar, antiferromagnetic and paramagnetic phases, respectively. Temperature dependences of (b) resistivity $\rho(T)$ and magnetic susceptibility $\chi(T) = M/H(T, H_0)$ and (c) thermal conductivity $\kappa(T, H_0)$ and specific heat $C(T, H_0)$ for various direction and intensity of applied magnetic field. $T_N$ and $T_Q$ denote the temperatures of two magnetic phase transition.

However, it is not possible to ignore experimental facts concerning the effects of electron phase separation [38−39], electron nematicity [46], FM instability [47−52] as well as the spin polarons' scenario of the charge transport proposed for $CeB_6$ in [52]. Therefore, it is of interest to examine in more detail the so−called AFQ phase (II in Fig. 1a) in this extraordinary HF metal. Thus, the goal of the present study is to clarify the nature of the AFQ phase in detailed magnetoresistance studies in combination with thermal conductivity and heat capacity measurements on high quality single domain crystals of $CeB_6$. In addition, precise XRD experiments are carried out at $T = 30$ K and 200 K (for comparison) to testify the electron density (ED) distribution at low temperatures. The obtained experimental results allow us to conclude in favor of the formation of spin polarons (ferrons) in phase II (AFQ) and to analyze the characteristics of these spin droplets. Finally, the paper proposes a theoretical model of the *inhomogeneous state with ferrons* in the $CeB_6$ matrix.

## 2. Experimental details

High quality $CeB_6$ single crystals were grown by vertical crucible-free inductive zone melting in argon gas atmosphere using a setup described in detail in [56−57]. The sample quality was characterized by XRD, microprobe and spectral analysis, magnetization and charge transport measurements. The single-domain crystals were cut from the same rods as in [32−33]. Heat capacity and thermal conductivity were measured in the temperature range 1.9−400 K in magnetic field up to 90 kOe (see Fig. 1c) on a commercial installation PPMS−9 (Quantum Design Inc.). Measurements of magnetoresistance (MR) $\rho(T, H)$ were performed in a four−terminal scheme with direct current (DC) commutation ($I = 1-100$ mA) for several crystals with current directions $\boldsymbol{I} \parallel$ [100], [110] and [111]. In the angular resolved studies of transverse MR performed in external magnetic field at temperatures in the 1.7−6 K range the step-by-step rotation of samples around the DC axis was used [58]. Small samples of $CeB_6$ of 0.05–0.3 mm in size were prepared for XRD data collection (see [58] for more detail). Single-crystal XRD data were collected at two temperatures 30 K and 200 K using a diffractometer XtaLAB Synergy−DW with a curved photon accumulation detector HyPix−Arc 150°. AgKα radiation ($\lambda = 0.56087$ Å) was applied in combination with N−Helix and Cobra Plus cryosystems (Oxford Cryosystems) with an open flow of nitrogen or helium gases. More information on the XRD experiment and the results of refining the structural model in the $Pm\bar{3}m$ symmetry group at temperatures of 30 K and 200 K are presented in Tables S1 and S2 in [59].

## 3. Experimental results.

**3.1. Magnetoresistance.** Fig. 2 shows the magnetic field dependences of resistivity $\rho(H, T_0)$ recorded for three principal directions of external magnetic field $\boldsymbol{H} \parallel [100]$, [110] and [111] in the temperature range 1.7−6 K, which corresponds mainly to the II (AFQ) phase of $CeB_6$ (see the $H-T$ diagram in Fig.1a). Step-like (near the Néel field $H_N(T)$) and knee-type (at $H_Q(T)$) anomalies on the $\rho(H, T_0)$ curves detected below $T_N \approx 2.3$ K and above $T_Q \approx 3.2$ K should be attributed to the III−II (AFM−AFQ) and II−I (AFQ−PM) phase transitions. Note that in phase II very strong (up to 95%) negative magnetoresistance (nMR) is observed, which usually is discussed in terms of field−induced suppression of the Kondo effect (see e.g. [1] and references therein). On the contrary, in [52] the nMR effect was interpreted in terms of the spin-polaron scenario of charge transport. The exponential behavior of resistivity $\rho(H, T_0)$ in phase II (Fig. 2) may be described by empirical formula

$$\rho(H, T_0) = \rho_0(T_0) + \rho_{\text{ferr}}(T_0) \cdot (T/T_N)^{2/3} \cdot \exp(-\mu_{\text{eff}(\rho)} H/k_B T_0) \qquad (1),$$

where $\rho_{\text{ferr}}(T)$ and $\rho_0(T)$ are field independent components, $k_B$ the Boltzmann constant and $\mu_{\text{eff}(\rho)}(T)$ denotes the magnetic field dependent activation energy of spin−polaron many body states, or in other words, the effective magnetic moments of spin droplets (ferrons) detected in the nMR study. The approximation of the nMR results by Eq. (1) is shown by solid lines in Fig. 2. The parameters $\rho_{\text{ferr}}$, $\rho_0$ and $\mu_{\text{eff}(\rho)}$ detected here are presented in Fig. 3 in comparison with the experimental resistivity $\rho(T, H=0)$ curve. It is seen in Fig. 3b, that $\mu_{\text{eff}(\rho)}(T)$ changes in the range 1.4−1.9 $\mu_B$ in phase II (AFQ) at temperatures 1.7−6 K, and moderate anisotropy (up to 0.15 $\mu_B$) of the effective magnetic moment is observed in the interval 2−3.5 K with maximal $\mu_{\text{eff}}$ values observed near $T_N$ for $\boldsymbol{H}//[100]$. The magnitudes of $\mu_{\text{eff}(\rho)}(T)$ are very similar to those from the ESR $g$ factor detected in the high frequency (~ 60 GHz) in ESR measurements of $CeB_6$ [49].

Taking into account the electron nematic effect in the magnetoresistance of $CeB_6$, which was observed in phase II (AFQ), reaching the maximal values of $\rho_{\boldsymbol{H}//[111]}/\rho_{\boldsymbol{H}//[100]}$ ~5−6% at 30 kOe [46], we verify a slight anisotropy of the charge transport parameters at $T_0=1.9$ K in MR measurements for different orientations of the external magnetic field varying both its intensity up to 80 kOe and the **H** direction in three different planes (100), (110) and (111).

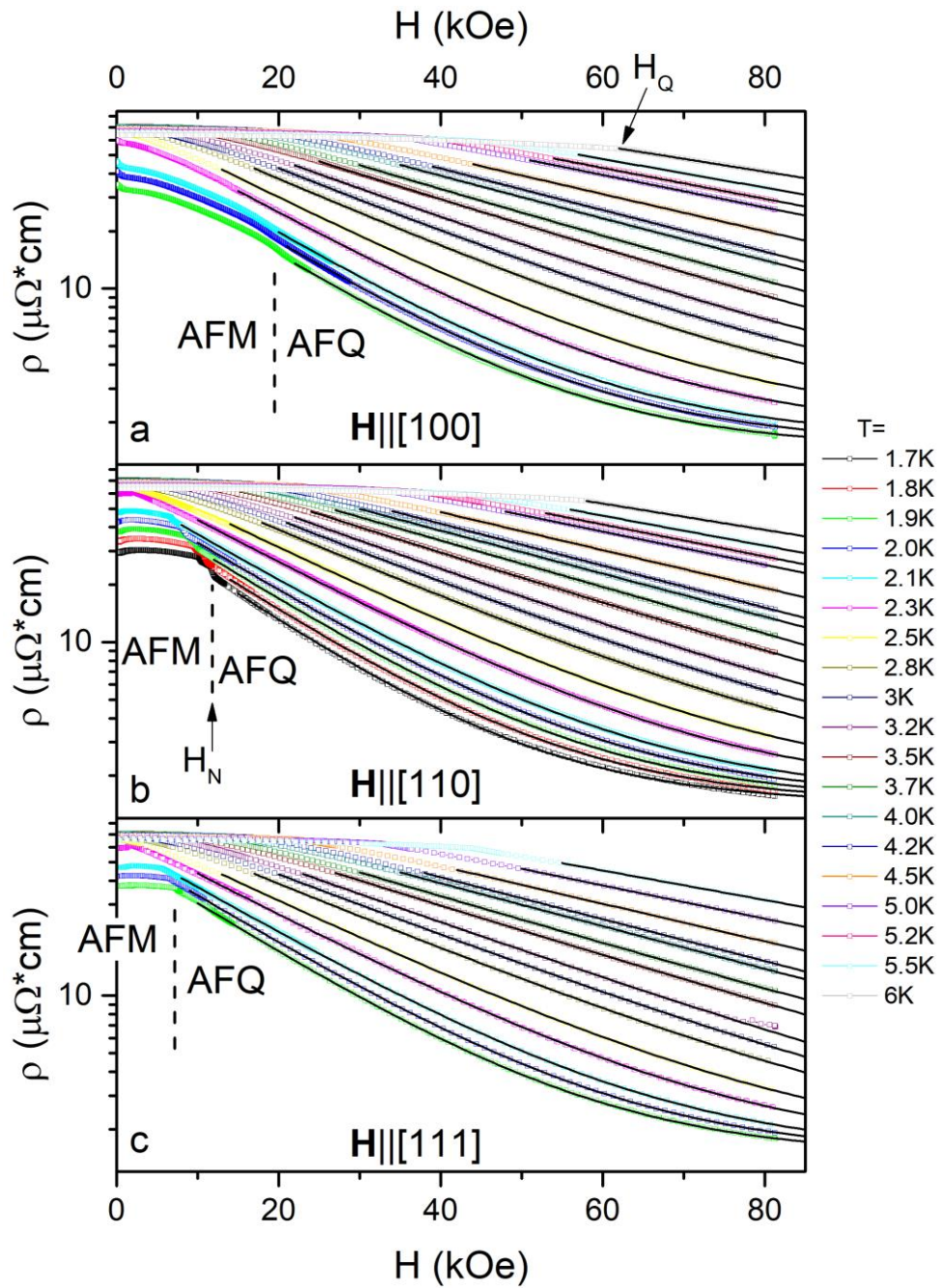

**Fig. 2.** Magnetic field dependences of resistivity at temperatures in the range 1.7−6 K for (a) **H**‖[100], (b) **H**‖[110] and (c) **H**‖[111]. AFQ and AFM denote the magnetic phases and solid lines show the fitting by Eq. (1).

In these experiments, the crystals of $CeB_6$ were rotated step−by−step around the DC direction (transverse MR configuration), and the field dependences of resistivity $\rho(H, \phi_0, T_0=1.9 K)$ were measured at fixed angles $\phi_0 \equiv \mathbf{n} \wedge \mathbf{H}$ (**n** is the normal vector to the lateral surface of the sample, see the sketch in the inset of Fig. 4a). Fig. 4 shows in the logarithmic plot the families of magnetoresistance $\rho - \rho_0 = f(H, \phi_0)$ curves recorded at $T_0=1.9$ K for various field directions in the planes **H**‖(100) (panel a), **H**‖(110) (b) and **H**‖(111) (c), demonstrating a good quality scaling of resistivity. The analysis of MR data in the AFQ(II) phase performed at $T_0 = 1.9$ K confirms only a slight anisotropy of both the resistivity components $\rho_{ferr}$, $\rho_0$ and effective moment $\mu_{eff(\rho)}$ detected in the wide range of magnetic field in the approximation given by Eq. (1) (Fig. 4).

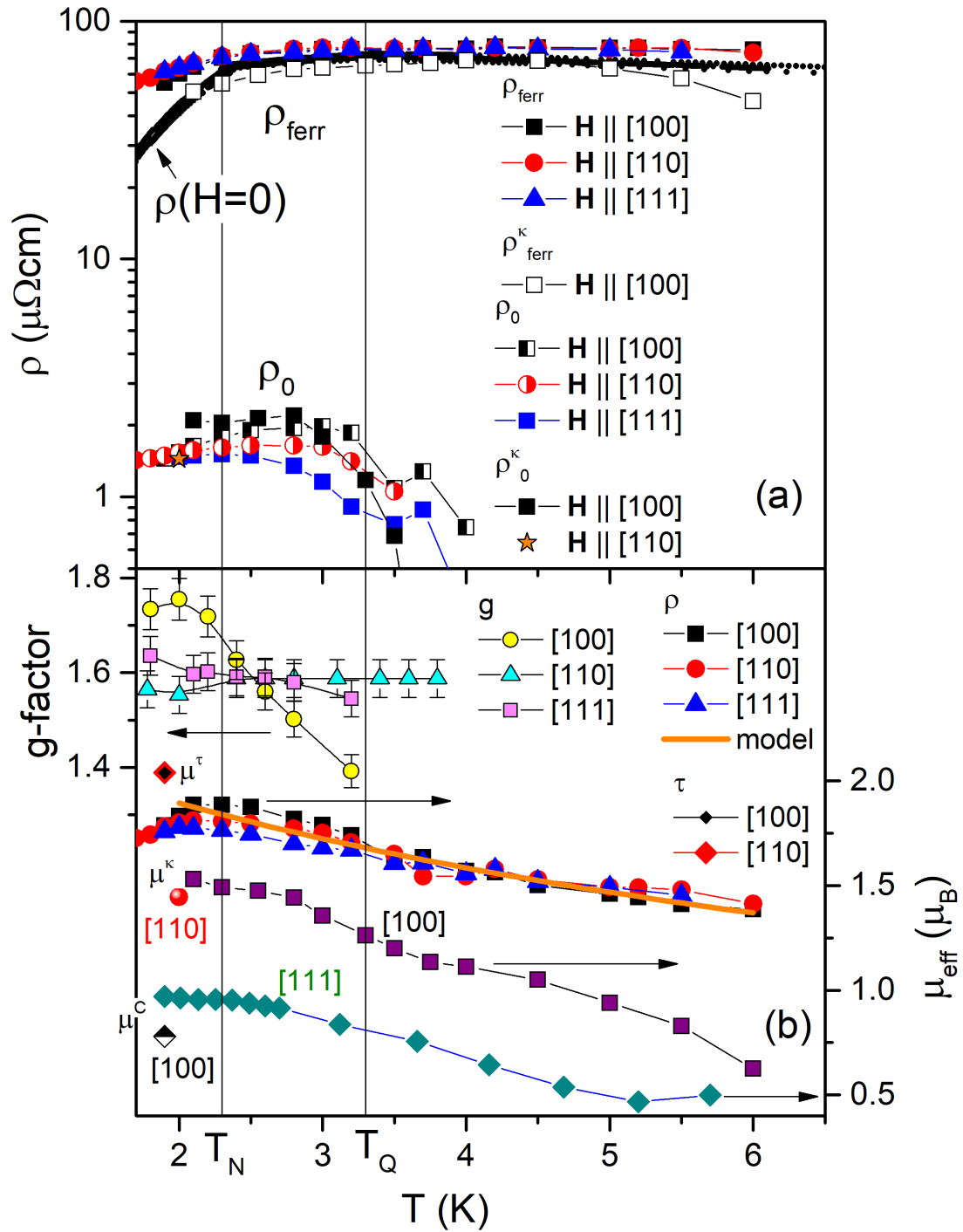

**Fig. 3.** Parameters (a) $\rho_{ferr}(T)$, $\rho_0(T)$, and (b) $\mu_{eff}(T)$ analyzed by the approximation of experimental results in framework of Eqs. (1–3, 6). $\rho_{ferr(\kappa)}(T)$, $\rho_{0(\kappa)}(T)$ are calculated using the Wiedemann−Franz relation $1/\rho_{i(\kappa)} = \kappa_i/L_0 T$ (see text for more detail). Solid (orange) line in (b) demonstrates the approximation by Eq. (13) with $T^* = 6$ K and $\alpha = 10$. The temperature dependences of ESR $g$ factor [49] are shown in panel (b) for comparison.

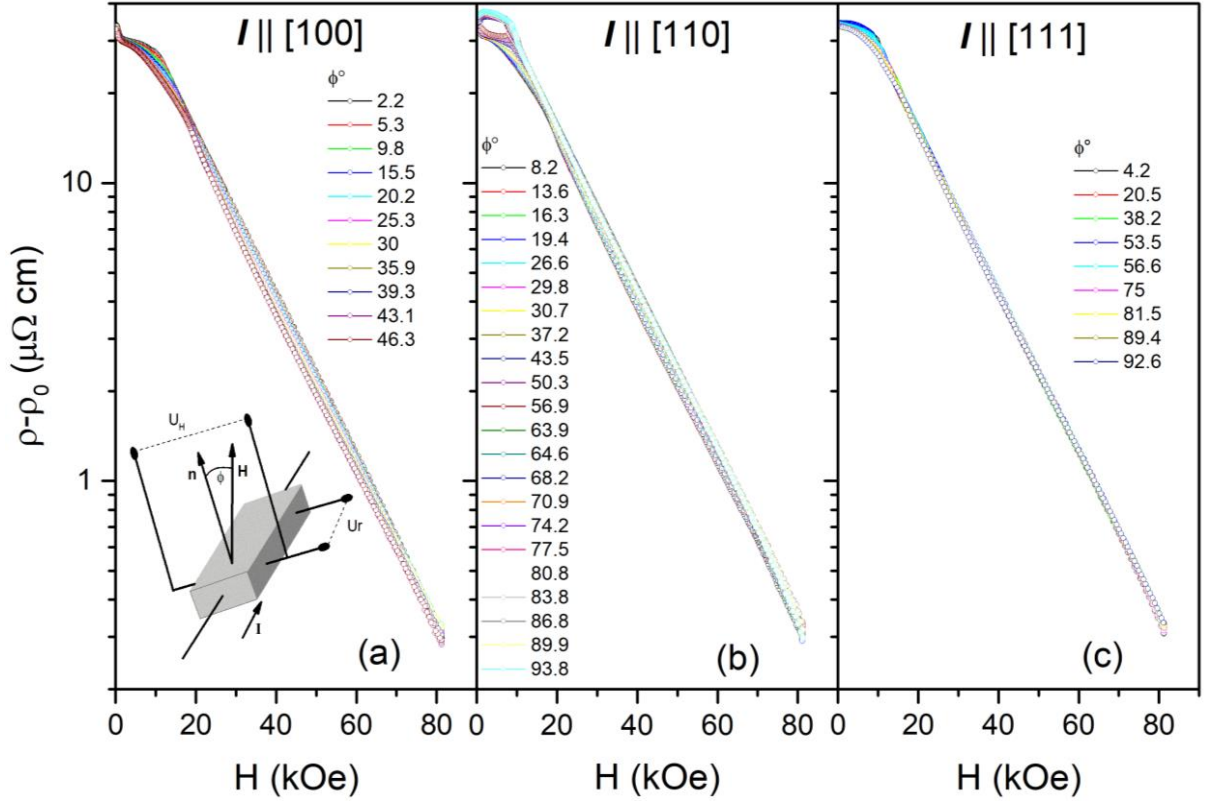

**Fig. 4.** Magnetic field dependences of the transverse magnetoresistance $\Delta\rho/\rho(H, \phi_0, T_0)$ at temperature $T_0 = 1.9$ K in magnetic field rotated step-by-step in planes (a) $\mathbf{H}\|(100)$, (b) $\mathbf{H}\|(011)$ and (c) $\mathbf{H}\|(111)$ (see text for more detail). AFQ and AFM denote the magnetic phases and $H_N$ is the Néel magnetic field. The sketch in the inset of panel (a) shows schematically the sample rotation.

**3.2. Thermal conductivity.** Fig.1c shows the temperature dependences of thermal conductivity $\kappa(T, H_0)$ of CeB$_6$ which were measured with a temperature gradient $\nabla T$ directed along the magnetic field in configuration $\nabla T\|\mathbf{H}\|[100]$. Very strong changes in external magnetic field of these temperature dependences $\kappa(T, H_0)$ were observed at low temperatures [60]. Here we concentrated on the precise studies of the magnetic field effects by recording $\kappa(H, T_0)$ curves in the range 1.9−6.5 K, which corresponds mostly to phase II (AFQ) (see Fig.1a). The data obtained (see Fig. 5a) allow us to conclude in favor of two contributions to the thermal conductivity, one of these components changing exponentially with the magnetic field. The results of Fig. 5a were approximated here by the formula

$$\kappa^{-1}(H, T_0) = \kappa_0^{-1}(T_0) + \kappa_{\text{ferr}}^{-1}(T_0) \exp(-\mu_{\text{eff}(\kappa)}H/k_B T_0), \quad (2)$$

with parameters $\kappa_0(T)$, $\kappa_{\text{ferr}}(T)$ (inset of Fig. 5) and $\mu_{\text{eff}(\kappa)}(T)$ (Fig. 3b) estimated by fitting. It is seen in Fig. 3b that $\mu_{\text{eff}(\kappa)}(T)$ changes monotonously in the range 0.6−1.5 $\mu_B$. The components $\kappa_0(T)$ and $\kappa_{\text{ferr}}(T)$ of thermal conductivity (inset of Fig. 5) decrease strongly with decreasing temperature. To establish a correspondence between $\rho_{\text{ferr}}(T)$ and $\rho_0(T)$, found in the MR experiment (see Eq. (1)), on the one hand, and $\kappa_{\text{ferr}}(T)$ and $\kappa_0(T)$ detected by approximation (2), the other, we use a simple Wiedemann−Franz relation $1/\rho_{i(\kappa)} = \kappa_i/L_0 T$ (where $L_0 = 24.5$ nW·Ω·K$^{-2}$ is Sommerfeld value of Lorentz number for the free electron gas and index i=0, ferr).

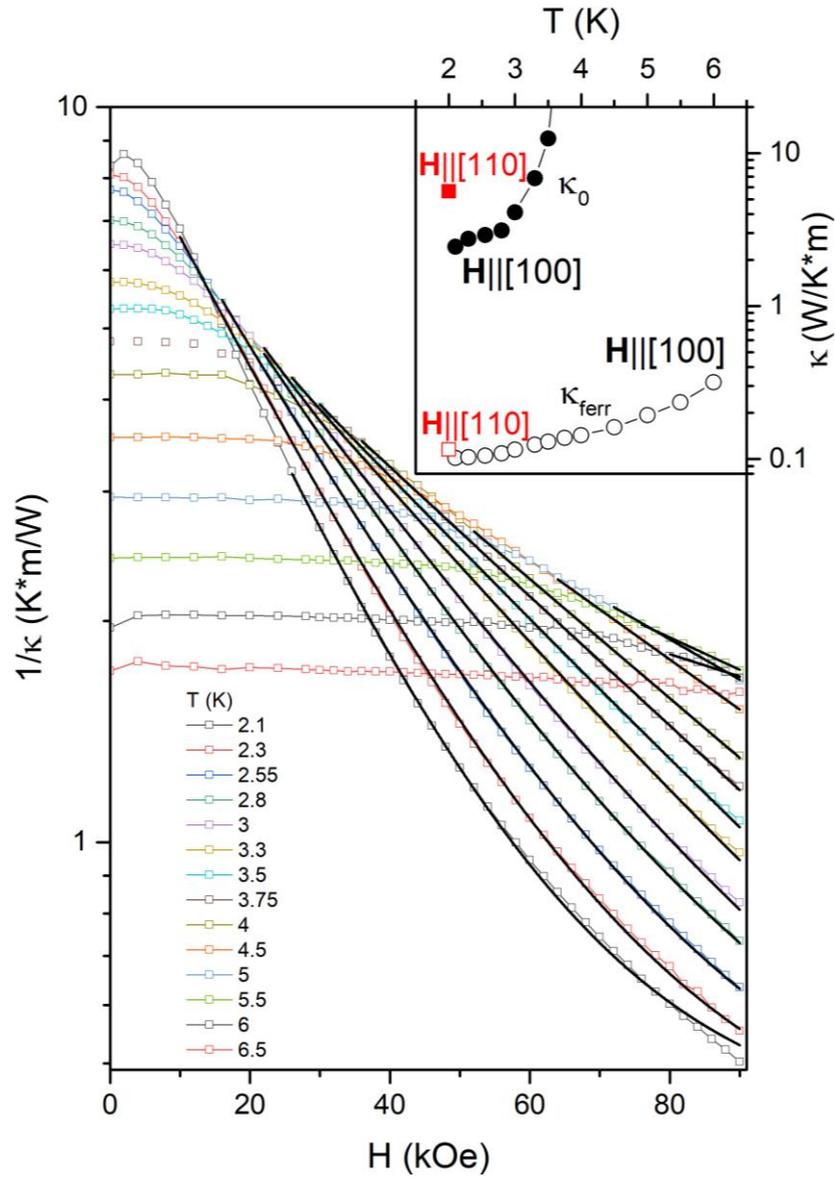

**Fig. 5.** Magnetic field dependences of the inverse thermal conductivity at temperatures in the range of 2.1−6.5 K in a magnetic field $\boldsymbol{H}\|[100]$. Thick solid lines show the fitting by Eq. (2) with parameters $\kappa_0(T)$ and $\kappa_{\text{ferr}}(T)$ presented in the inset and $\mu_{\text{eff}(\kappa)}(T)$ (see Fig. 3b).

The resistivity components $\rho_{0(\kappa)}(T)$ and $\rho_{\text{ferr}(\kappa)}(T)$ detected from the thermal conductivity data are shown in Fig. 3a together with parameters $\rho_{\text{ferr}}(T)$ and $\rho_0(T)$ estimated from the MR measurements testify a good correlation between these two sets of charge transport characteristics.

**3.3. Specific heat.** Fig. 1c demonstrates the specific heat $C(T, H_0)$ for CeB$_6$ recorded at $H_0$=0, 20, 60 and 90 kOe with a strong dependence on magnetic field at helium temperatures. The locations of magnetic phase transitions at $T_N$ and $T_Q$ are determined by magnetic field (see Fig. S1 in [59]), and the data allow us to refine the magnetic $H-T$ phase diagram (Fig. 1a) for the studied CeB$_6$ crystals. Fig. 6a shows the magnetic field dependences $C(H, T_0)$ in the range 1.9−6 K for the field direction $\boldsymbol{H}\|[111]$. To investigate the possible field induced anisotropy of the heat capacity we have measured also $C(H, T_0)$ at $T_0 = 1.9$ K for $\boldsymbol{H}\|[100]$ and $\boldsymbol{H}\|[110]$, these results are presented in Fig. 6b for comparison.

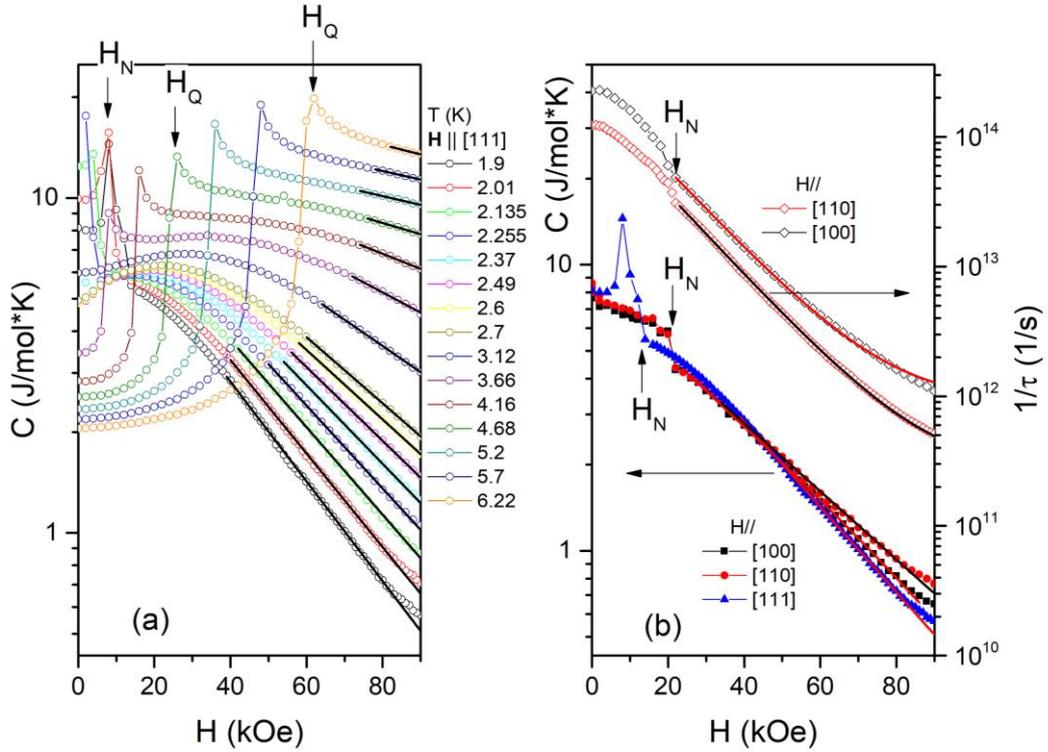

**Fig. 6.** Magnetic field dependences of the specific heat $C(H, T_0)$ (a) at temperatures in the range of 1.9−6.2 K in a magnetic field $\boldsymbol{H}//[111]$ and (b) at $T=1.9$ for $\boldsymbol{H}//[100]$, $\boldsymbol{H}//[110]$ and $\boldsymbol{H}//[111]$. Thick solid lines show the fitting by Eq. (3). Panel (b) presents the estimates of the relaxation rate by Eq. (4).

The analysis of the specific heat behavior in the II(AFQ) phase was made using relation

$$C(H, T_0) = C_1(T_0)\exp(-\mu_{\text{eff}(C)}H/k_B T_0) + C_0, \tag{3}$$

and the deduced $\mu_{\text{eff}(C)}(T)$ is compared with values $\mu_{\text{eff}(\rho)}$ and $\mu_{\text{eff}(\kappa)}$ in Fig. 3b.

**3.4. XRD studies.** As recently found, electron and lattice instabilities develop in CeB$_6$ being important factors, which modify the properties of this archetypal HF compound with magnetic clusters of Ce ions. Both dynamic charge stripes and sub-structural charge density waves (s-CDW) were detected in the precise XRD studies of CeB$_6$ in the temperature range 85−500K [38−39], and electron phase transition induced by changes in the stripe configurations were observed at $T_c \sim 340$ K [39]. Taking into account that the weak localization regime of charge transport preceds the transition into the so-called AFQ (II) phase of CeB$_6$, it is important to investigate the variation of the stripe patterns and the s-SDW on cooling. To elucidate the nature of changes in crystalline and electron structure in the $T \leq 200$ K range, we have carried out precise X−ray diffraction measurements at two temperatures, $T_0 = 200$ K, well above the interval of weak localization of charge charriers (Fig. 1b and [38]), and $T_0 = 30$ K, the lowest temperature achieved in our XRD facility. The structure of the single crystals was refined in the $Pm\bar{3}m$ symmetry group (Fig. 7a−7c). The main characteristics of the XRD experiment and the results of structural model refinement are presented in Tables S1 and S2 in [59]. An independent estimate of the electron density (ED) distribution in the crystal is obtained using the maximum entropy method (MEM) without involving symmetry restrictions. The MEM maps of ED in {100} and {110} planes are presented in Figs. 7d, 7e and 7f (see Fig. S2 in [59] for more detail).

The comparison of the structural characteristics and the MEM maps of ED allow us to come to the following conclusions:

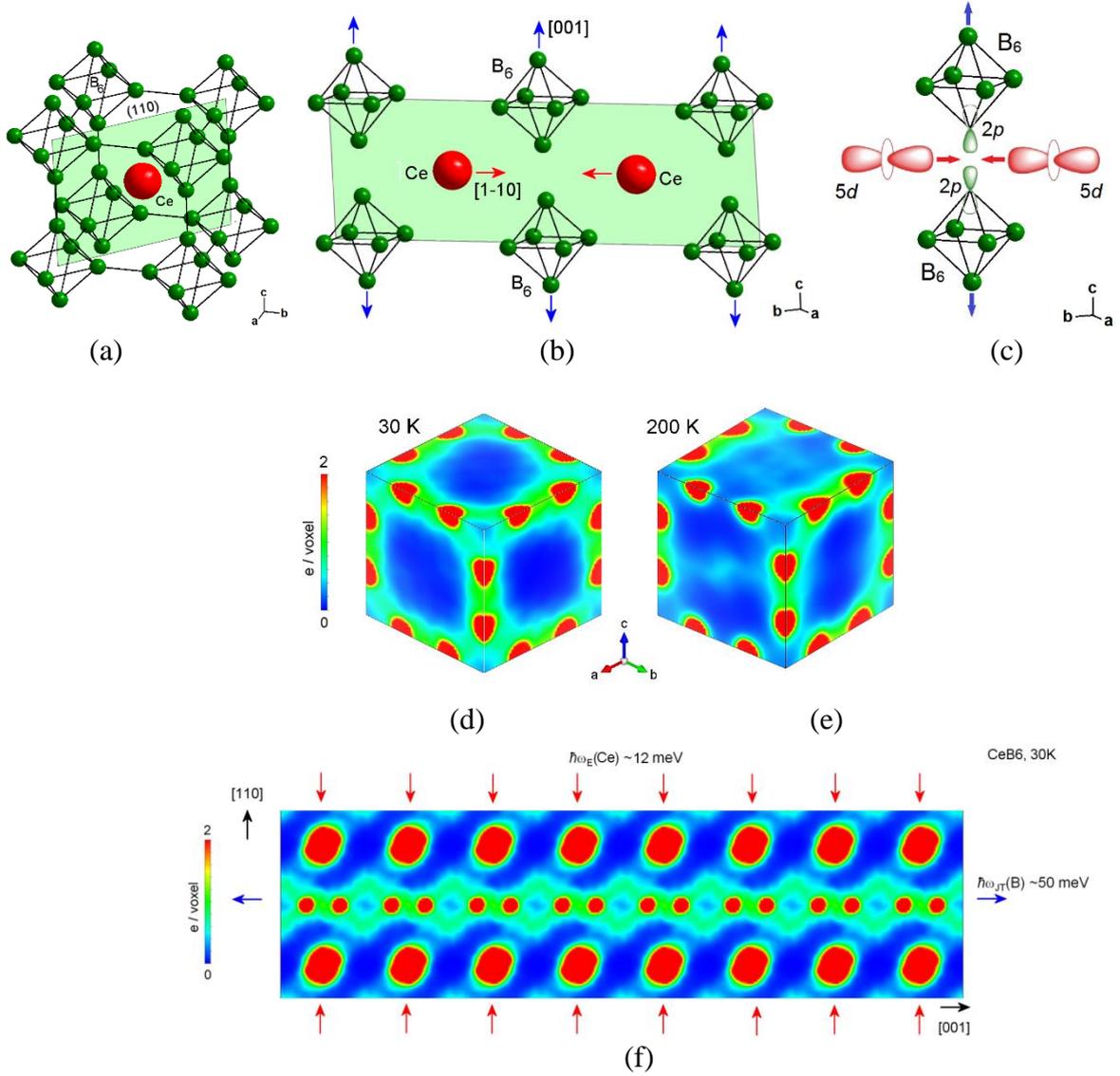

**Fig. 7.** (Color online) (a) Crystal structure of CeB$_6$. (b) Crystal structure in the projection on the (110) plane (this plane is highlighted in green in panels (a) and (b)), blue arrows indicate the vibrations of B$_6$ clusters (the collective JT mode [61−62]) in the <100> direction giving rise to quasi−local vibrations (the Einstein modes [63−64], shown by red arrows) of Ce ion pairs in one of the directions of the <110> family and to periodic changes of $5d-2p$ hybridization of the band states (see panel (c)). In panels (d)−(e), we show the electron density distribution in {100} planes obtained by the maximum entropy method (MEM) at: (d) $T = 30$ K and (e) $T = 200$ K. (f) MEM−constructed map of the ED with a dynamic charge stripe in one of the {110} planes and the JT boron ($\hbar\omega_{JT} \sim 50$ meV) and Einstein cerium ($\hbar\omega_E \sim 12$ meV) modes forming this dynamic stripe (see text). ED peaks were cut off at a height of 2 e/voxel to highlight fine details of the ED distribution in the lattice interstices.

(1) CeB$_6$ exhibits small static JT distortions of the cubic lattice (about 0.003 Å in the linear and up to 0.04° in angular sizes, see Table S2 in [59]), which do not require a transition to a noncubic structural model. However, the cooperative dynamic structural instability of the boron frame (ferrodistortive JT effect) leads to a modulation of the hybridization of the $5d$(Ce) and $2p$(B) conduction band states [38, 39, 65] (see schematics in Figs. 7b and 7c), which is the cause for the differences in the distribution of interstitial ED in symmetry−equivalent {110} planes (see also Fig. S2 in [59]);

(2) the ED maps at Ce sites located within symmetry−equivalent {110} planes also exhibit differences (noticeable deviations from the spherical shape, see Fig. S2 in [59]) suggesting the formation of vibrationally bound clusters of rare-earth ions established previously for different $RB_6$ compounds [38, 39, 65]. Moreover, at $T = 200$ K a s-CDW is observed in (1−10) and (10−1) planes, and a stripe arranged on $5d$(Ce) and $2p$(B) states is detected only in the (01−1) plane (see Fig. S2 in [59]);

(3) the most significant differences in the ED distribution at $T = 200$ K and 30 K are observed in the {100} planes. They are associated with the formation at $T = 30$ K of a network of dynamic charge stripes along the edges of the cubic cell (Fig. 7d), instead of the segments of charge stripes located at 200 K in some of the B−B bridges between neighboring $B_6$ clusters (Fig. 7e). The ED segments at the B−B bridges seem to be related to two factors: to electrons located at the B−B bond and to quasilocal vibrations ($\hbar\omega_E \sim 12$ meV [63−64]) of Ce−Ce pairs along the <110> directions, leading to a periodic change of the $5d$–$2p$ hybridization (see the scheme in Figs. 7b and 7c).

Further, when the temperature decreases from 200 K to 30 K, extended dynamic stripes are formed predominantly by the $2p$ orbitals of boron involving also the $5d$ states of cerium along the <110> and <111> directions in $CeB_6$ (Figs. 7d and 7f). It is reasonable to assume that these filamentary three-dimensional structures of fluctuating charges are related to the collective JT mode arising in the rigid boron cages (see the scheme in Fig. 7f). This collective mode with an energy $\hbar\omega_{JT} \sim 50$ meV ~ 580 K reliably manifests itself in optical conductivity measurements [61] and in the Raman spectra of $LaB_6$ [62].

### 4. Discussion

**4.1.** *Homogeneous* **(single-phase) state description.** Even though the studies of the complex phase diagram of the heavy fermion system $CeB_6$ started more than 55 year ago [66], this unusual antiferromagnetic metal with a cubic lattice has been the subject of continued experimental and theoretical investigations up to present days, and especially the nature of phase II (AFQ, Fig.1a) is still under debates (see e.g. [45−52, 56]). Indeed, the order parameter responsible for phase II proved to be difficult to determine, since it is 'hidden' to neutron diffraction measurements in zero field [67]. It is stabilized by an applied magnetic field above 1 kOe [68, 69], as evidenced by the rapidly enhanced transition temperature $T_Q(H)$ in low field [45−52, 56] (Fig. 1a), and it is believed that in the field dipolar moments at the $Ce^{3+}$ sites are induced, resulting to magnetic order with a wave vector $\mathbf{k}_0 = 2\pi/d$ [½,½,½] ($d \approx 4.14$ Å is the lattice constant, see Table S1 in [59]) observed by polarized neutron [70−71] and resonant X−ray [72] scattering. It was found in [71], that the intensity of the magnetic Bragg reflection is *two orders of magnitude weaker* than those due to the basic magnetic structure. The peak has a width of the other Bragg reflection observed in the AFM phase (III in Fig. 1a), but widens abruptly at $T = T_N$ with a simultaneous increase of intensity in phase II [71]. The correlation length of the antiferromagnetic fluctuations in phase II just above $T_N$ is of the order of 10 Å. The peak intensity decreases to zero upon warming up to $T \sim 7$ K with no visible anomaly at the antiferroquadrupolar ordering temperature $T_Q \approx 3.3$ K. Plakhty *et al.* note [71] that the features of the magnetic peak [½,½,½] are typical for *the itinerant magnetism*, but later on this result was explained by an ordering of the quadrupolar moments $Q$ and $−Q$ corresponding to the splitting into two doublets of $\Gamma_8$ ground state of $^2F_{5/2}$ multiplet of $Ce^{3+}$ ion. Note, that antiferromagnetic ordering in phase II under magnetic field has also been observed by $^{11}$B NMR [73, 74], but with a different, triple-**k** structure: $\mathbf{k}_1 = 2\pi/d$ [½,0,0], $\mathbf{k}_2 = 2\pi/d$ [0,½,0], $\mathbf{k}_3 = 2\pi/d$ [0,0,½]. Furthermore, the result of a $\mu$SR study [75] cannot be explained in terms of the field−induced antiferromagnetic structure proposed either in the neutron diffraction or in the NMR studies. Besides, it has been claimed by Kasuya *et al.* [75, 76] that there is no quadrupolar ordering in the phase II. According to [75, 76], single-site dynamical and pair Jahn–Teller distortions explain the hidden ordering that is responsible for the field−induced antiferromagnetism. In other words this means some antiparallel displacements of the Ce atoms along the [001]−type direction with an ordering wave vector $\mathbf{k}_0 = 2\pi/d$ [½,½,½]. Distortions of the CsCl type structure were predicted also by Lovesey [77], but it was argued in [77] that the tetragonal deformation transforms only the $B_6$ octahedrons preserving the cubic Ce sub−lattice in $CeB_6$. In [78], the spin density distribution of $CeB_6$ was obtained by the maximal entropy method using polarized neutron diffraction data collected at $H = 59$ kOe and $T = 1.6$ K (phase II, Fig.1a). The authors of [78] have revealed the presence of significant amounts of localized spin moment at non-atomic sites between the nearest-neighbor boron atoms at the center of the triangle surface in the $B_6$ octahedron network. On the contrary, similar experiments at $H=40$ kOe and $T=4.2$ K, as well as a MEM analysis of the data was undertaken in [79] which allowed to conclude that the magnetization is localized on cerium sites only. Subsequently, developing the AFQ scenario of I−II phase transition and resolving the disagreement between the results of neutron scattering and NMR investigations on $CeB_6$, a significant efforts were undertaken by a number of theorists, and along with the ordering of $O_{xy}$ quadrupoles in applied field $H\|<001>$ also the $T_{xyz}$ AF octupolar interaction was included in the new model (see review [80] and references therein for more detail).

Simultaneously with the commonly used AFQ approach, the two-jority of Ce−based dense Kondo systems, the Hall coefficient $R_H$ is both negative and nearly independent on temperature and magnetic field in the paramagnetic (PM) phase of $CeB_6$ (phase I in Fig.1a) between 7 − 300 K [52], which goes against the predictions of the skew-scattering models for dense Kondo systems [83−84]. An alternative explanation was proposed by Sluchanko *et al.* [52] in the charge transport and magnetization study. In their scenario the formation and subsequent increase upon cooling of the spin-polaron resonance in vicinity of the Fermi level induces the rearrangement of many-body states in the temperature range $T_Q < T < 7$ K. If the Stoner-type criterion holds, the spin−polaron states give rise to formation of nanosize ferromagnetic domains (ferrons) with a localization radius $a_{SP} \approx 5$ Å, being responsible for the activation dependences of the Hall coefficient $R_H$ and magnetic susceptibility $\chi_p$ of the type $\chi_P(T) \sim R_H(T) \sim exp(E_{SP}/k_BT)$, where the activation energy $E_{SP}/k_B \approx 3.3$ K $\approx T_Q$ was associated with the bound energy of the spin-polarized many-body states. Thus, in the framework of this approach, a phase transition at $T_Q$ occurs in the system of interacting ferrons, which are SDW antinodes [52, 85], that frees the local moments of $Ce^{3+}$ ions and leads to the observed enhancement of the magnetic response.

**4.2. JT lattice instability and electron phase separation (stripes).** All the models listed in 4.1 describe *the homogeneous single−phase state*, which is not valid in the case of $CeB_6$. Also the results of resistivity $\rho(T)$ and magnetization $M(T)$ measurements obtained very recently [38−39] at temperatures 4−800 K allow to conclude that both the charge transport and the magnetic properties of the archetypal heavy fermion $CeB_6$ compound cannot be described by the single−ion Kondo lattice model. Indeed, instead of the Currie−Weiss type behavior a single power−law dependence of magnetization $M(T) \sim (T-T_C^{rand})^{-0.8}$ was detected in the wide range 5−800 K for three principal directions of magnetic field up to 50 kOe (Fig. 1b), indicating the Griffiths phase scenario in $CeB_6$ above $T_C^{rand} = T_Q \approx 3.3$ K. In addition, the power-law behavior of the magnetic contribution to resistivity $\rho_m(T) \sim T^{-0.4}$ was deduced in [38] in the range 8–90 K (see also Fig. 1b), far above the Kondo temperature $T_K \sim 1$ K and well below the crystal electric field splitting $\Delta_{CEF} \approx 530$ K, proposed for the ground state $^2F_{5/2}$ multiplet, attributing the regime of weak localization of charge carriers in the nanosize clusters of Ce ions. Moreover, precise XRD experiments in the range 85−500 K established small *static* Jahn−Teller distortions of the simple cubic lattice, which were accompanied by the appearance of (*i*) *dynamic charge stripes* along selected <110>, <100>, and <111> directions in combination with (*ii*) *vibrationally coupled pairs* of Ce ions in $CeB_6$ crystals [38−39]. Fourier maps of ED confirm the formation of nanosize magnetic clusters of Ce ions (Griffiths phase) in $CeB_6$, which allows a new *heterogeneous* approach to interpret the properties of this archetypal magnetic heavy fermion hexaboride [38−39].

The ED maps obtained in the present study at $T = 30$ K (Fig. 7 and Fig. S2 in [59]) show a dramatic transformation of charge stripe patterns upon cooling within a weak localization regime.

As a result, three−dimensional filamentary structures of fluctuating charges are located strictly along the $B_6$ clusters, corresponding to the $2p$ states of boron participating in the collective JT mode with an energy of $\hbar\omega_{JT} \sim 50$ meV $\sim 580$ K (Fig. 7f). The boron JT cooperative dynamics induces quasilocal vibrations ($\hbar\omega_E \sim 12$ meV [63, 64]) of Ce−Ce pairs along the <110> and <111> directions (see the scheme in Fig. 7f and Fig. S2 in [59]). The contribution of the $5d$ states of Ce in the $2p$ type stripes of boron chains is detected clearly both in Fig. 7f and Fig. S2 (planes (101) and (01−1) at $T = 30$ K in [59]). We propose that these transversal to stripes of quasilocal vibrations of Ce pairs are responsible for the magnetization of $5d$ states in the stripes by $4f$ magnetic orbitals of Ce ions forming FM spin droplets (ferrons) in the nearest vicinity of dynamic charge stripes in $CeB_6$. In the nonmagnetic $LaB_6$ a similar mechanism of localized superconductivity induced by the Cooper pairing mediated by quasi−local vibrations of La ions was proposed very recently [86]. In our opinion, these ferrons magnetized by $4f$−$5d$ spin fluctuations and arranged between two Ce sites along <110> and <111> directions (see Fig. 7f and Fig. S2 in [59]), may be the cause of the discussed above anomalies of the charge transport and thermodynamic characteristics of $CeB_6$. In this scenario, one can assume a ferromagnetic arrangement of the magnetic moments of two vibrationally coupled Ce ions, which, because of the induced $4f$−$5d$ spin fluctuations, is accompanied by FM polarization of the $5d$ states within the cerium pair. As a result, the size of the FM spin droplet (ferron) may be estimated as $d \cdot \sqrt{2}$, $d \cdot \sqrt{3} = 6$−$7$ Å.

### 4.3. Microscopic parameters and phenomenological model of ferrons
**4.3.1. Estimations.** To evaluate roughly both the relaxation time and concentration of Drude type electrons and non-equilibrium (participating in the Jahn−Teller collective mode) charge carriers, we used simple relations, which, strictly speaking, are valid only for Drude electrons

$$\kappa(H, T_0) = \tfrac{1}{3} C(H, T_0) \cdot v_F^2 \cdot \tau(H, T_0) \quad (4)$$

and

$$\rho_i^{-1}(H, T_0) = n_i \cdot e^2 \cdot \tau_i(H, T_0)/m_i^*, \quad (5)$$

where $v_F$ is Fermi velocity, $\tau$ the relaxation time, $m_i^*$ the effective mass of the charge carriers with index $i = 0$, ferr and $e$ the charge of electron. Using $v_F \approx 1.1 \cdot 10^6$ sm/s detected in studies of magneto−acoustic quantum oscillations in $CeB_6$ for $\boldsymbol{H}\|[100]$ in the interval $H = 63$−$83$ kOe [87] we plot in Fig. 6b the field dependences of the relaxation rate obtained from Eq. (4) at $T_0 = 1.9$ K. Near exponential dependence of relaxation rate was approximated by relation

$$\tau^{-1}(H, T_0) = \tau_{\text{ferr}}^{-1}(T_0) \exp(-\mu_{\text{eff}(\tau)} H/k_B T_0) + \tau_0^{-1}(T_0), \quad (6),$$

where $\tau_{\text{ferr}}^{-1}(T_0) \sim 7.7 \cdot 10^{13}$ s$^{-1}$ and $\tau_0^{-1}(T_0) \sim 3 \cdot 10^{11}$ s$^{-1}$ were estimated from fitting. Then, taking, because of the local character of charge and spin fluctuations, $m_{\text{ferr}}^* \approx m_0$ for the effective mass of electron in ferrons and $m_0^* \approx 6.5\, m_0$ [87] for the other conduction electrons in $CeB_6$, we deduce from Eq. (5) $n_{\text{ferr}} \approx 0.6 \cdot 10^{22}$ cm$^{-3}$ and $n_0 \approx 0.5 \cdot 10^{22}$ cm$^{-3}$. It is worth noting, that the rough estimate of the charge carriers' concentration $n \approx n_0 + n_{\text{ferr}} \sim 1.1 \cdot 10^{22}$ cm$^{-3}$ in $CeB_6$ is related closely to the value $n \approx 1.2 \div 1.5 \cdot 10^{22}$ cm$^{-3}$ obtained from Hall effect and optical studies of both the non−magnetic reference compounds $LaB_6$ [61, 88] and $YB_6$ [89, 90], and also deduced in the paramagnetic phase of $CeB_6$ [52, 91], $PrB_6$ and $NdB_6$ [91], and substitutional solid solutions $Gd_xLa_{1-x}B_6$ [88]. Moreover, the relaxation rate $\tau_0^{-1}(T_0) \sim 3 \cdot 10^{11}$ s$^{-1}$ obtained in present study is very similar to the frequency of dynamic charge stripes $\nu_s \sim 240$ GHz detected in optical measurements of the narrow−gap SCES $Tm_{1-x}Yb_xB_{12}$ [43−44, 92]. Note also that similar values of the spin fluctuation rate of $4f$ electrons $2$−$4 \cdot 10^{11}$ s$^{-1}$ were deduced just above $T_N$ in the $\mu$SR studies of $CeB_6$, in phase II (AFQ) [93]. We need to suppose in our considerations that the carriers' scattering on charge and spin fluctuations in dynamic stripes is one of the two dominant mechanisms for conduction electrons in $CeB_6$. Let us point out also, that in hexaborides $Gd_xLa_{1-x}B_6$ only a small amount of itinerant electrons behave as Drude type charge carriers while 50−70 % of charge carriers are involved in collective oscillations of the electron density coupled to vibrations of both the Jahn−Teller unstable rigid boron cage and rattling modes of heavy RE ions loosely bound to the lattice [61,88]. Besides, the relaxation rate $\tau_{\mathrm{ferr}}^{-1}(T_0) \sim 7.7 \cdot 10^{13}$ s$^{-1}$ obtained here from the analysis by Eq.(4) of thermal conductivity and heat capacity data in CeB$_6$ may be compared with $\tau_{\mathrm{JT}}^{-1}$(YB$_6$) $\sim 12.6 \cdot 10^{13}$ s$^{-1}$ [90], $\tau_{\mathrm{JT}}^{-1}$(LaB$_6$) $\sim 6.9 \cdot 10^{13}$ s$^{-1}$ [61, 88] and $\tau_{\mathrm{JT}}^{-1}$(GdB$_6$) $\sim 10.7 \cdot 10^{13}$ s$^{-1}$ [88] values, deduced in optical studies from relation $\tau_{\mathrm{JT}}^{-1} = 2\pi\gamma_{\mathrm{peak}}$, where $\gamma_{\mathrm{peak}}$ is the damping of the collective mode, for non−equilibrium charge carriers participated in collective modes in the YB$_6$, LaB$_6$ and GdB$_6$, respectively. Note also the relaxation rate $\tau^{-1}$(YbB$_6$) $\sim 13 \cdot 10^{13}$ s$^{-1}$ deduced from Hall effect studies of YbB$_6$ [94].

The obtained magnetic moment of ferrons $\mu_{\mathrm{eff}}(T) = 1.4-1.9$ $\mu_B$ (Fig. 3b) may be put into agreement with the ESR data in the following way. For g factor $g = 1.4-1.8$ (Fig. 3b) and quantum number $J_{\mathrm{eff}} = 1$ the magnitude of the magnetic dipole will be $\mu = 1.4 \div 1.8 \mu_B$. The case $J_{\mathrm{eff}} = 1$ corresponds here to the FM arrangement of two vibrationally coupled magnetic moments of Ce ions, which should be taken into account in the combination with FM spin polarized 5$d$ states within a cerium pair. Just these complex Ce−Ce pairs are responsible for the static transport and thermodynamic properties, although the coupling into Ce pairs may be broken at frequencies ~60 GHz (~0.25 meV~ $T_Q$) and higher, and thus an "individual" ESR of Ce ions with $J_{\mathrm{eff}}=1/2$ develops. In this approach the magnitude of $\mu_{\mathrm{eff}(\rho)}(T)$ is approximately twice higher than the magnetic moment $\mu = g\mu_B J_{\mathrm{eff}} \sim 0.8\mu_B$ following from the g factor value found in high frequency ESR experiments. It is worth noting that a very close value $g = 1.9 \pm 0.07$ was estimated for the magnetic peak at R(½ ½ ½) point in the Brillouin zone from inelastic neutron scattering experiments with a neutron energy of $E = 3.5$ meV [95]. Theory of magnetic resonance in HF metals developed in [96−98] explains the ESR phenomenon as a coupled mode of itinerant electrons and localized magnetic moments moving at the same frequency. For that reason, we assume that the proposed explanation does not contradict to the ferron model considered in the next section.

**4.3.2. Phenomenological model of ferrons.** In the analysis of experimental data, let us assume that we are dealing with two types of charge carriers. The first type corresponds to those filling the conduction band. Such charge carriers with concentration $n_0$ do not contribute to the observed strong magnetic field dependence of the specific heat $C$, thermal conductivity $\kappa$, and resistivity $\rho$. Charge carriers of the second type are located within charge inhomogeneities (ferrons) and contribute to the transport and thermal characteristics via charge transfer between the inhomogeneities. The properties of ferrons appear to be highly sensitive to the applied magnetic fields, and this feature can lead to pronounced magnetic field dependences of $C$, $\kappa$, and $\rho$. Therefore, we put in the next part the main emphasis on the charge carries related to ferrons.

We assume that the number of ferrons is constant and approximately equal to the number of itinerant charge carriers of the second type ($n_{\mathrm{ferr}}$). The experimental results described above allow us to argue that each ferron contains in the equilibrium state one electron of the second type. Following Refs. [99−101], we also assume that all spins within a ferron are parallel to each other, but are deviated by an angle $\theta$ from the direction of the applied magnetic field. We neglect the interaction between ferrons assuming that their density is not high.

To describe the experimentally observed $\rho(H)$, $\kappa(H)$, and $C(H)$ curves (Figs. 2, 4−6) it is necessary to estimate the ferron size and energy. For simplicity, we neglect the small anisotropy of the system and treat ferrons as spherical particles with a radius $R$. Correspondingly, following the approach of Ref. [100], we can write the $R$-dependent part of the free energy of the ferron in the form

$$F = \frac{\pi^2 t d^2}{R^2} + \frac{4\pi R^3}{3d^3}[JzS^2 + k_B T\ln(2S + 1) - \mu_B gSH \cos \theta]. \qquad (7)$$

Here, the first term in Eq. (7) is the part of kinetic energy of an electron within the ferron related to the size quantization, $t$ is the hopping amplitude for an electron inside the ferron, and $d$ is the lattice constant. Terms in the square brackets are, respectively, the Heisenberg energy of the local spins $S$, which are aligned in the ferron due to the exchange interaction, the entropy contribution, and the energy related to the interaction of local spins with the applied magnetic field $H$. Here, $J$ is the constant of AFM exchange interaction between the atoms in Ce−Ce pairs forming a ferron along <110> and <111> directions, $z = 12$ is the number of next-nearest magnetic ions located in the six {110} planes in the lattice, $\mu_B$ is the Bohr magneton, and $g$ is the Landé $g$ factor. The minimization of $F$ with respect to $R$ provides the values of the radius of an equilibrium single−electron ferron and of its free energy

$$R(T,H) = \frac{R(T,0)}{(1 - bH\cos\theta)^{1/5}}, \quad F_1(T,H) = \frac{5\pi^2 t d^2}{3R^2(T,H)}, \tag{8}$$

where $R(T,0)$ is the ferron radius at $H = 0$

$$R(T,0) = d\left[\frac{\pi t}{2k_B(T + T^*)\ln(2S+1)}\right]^{\frac{1}{5}}, \quad k_B T^* = \frac{JzS^2}{\ln(2S+1)},$$

$$b = \frac{\mu_B g S}{k_B(T + T^*)\ln(2S+1)}. \tag{9}$$

The electron hopping gives rise to the formation of two−electron and empty droplets. Following Refs. [100−101], we can neglect the existence of excited states within ferrons and assume that the decay time of an empty droplet is much larger than the characteristic hopping time. In the empty droplet, we should omit the electron kinetic energy. As a result, the free energy of an empty droplet is $F_0 = 2\pi^2 t d^2/3R^2(T,H)$. In the two−electron ferron, it is necessary to take into account the Coulomb repulsion of extra electrons and add the corresponding term to the free energy, $F_2 = F_1 + e^2/\varepsilon R^*(T,H)$, where $\varepsilon$ is the dielectric constant and $R^*(T,H)$ is the average distance between two electrons in the droplet. It is reasonable to assume that this value is close to the diameter of the ferron. In this case, we can rewrite the free energy of the two−electron droplet as $F_2 = F_1 + e^2/\varepsilon\gamma R(T,H)$, where $\gamma \approx 2$. Note, that $F_2 \gg F_{1,0}$ since we have $e^2/d \gg td/R$ in any realistic case. Let the total number of ferrons be $N$, and $N_1$ is the number of the single−electron ferrons, whereas $N_2 = N_0$ are the numbers of two−electron and empty ferrons, respectively ($N = N_1+N_0+N_2$). We are interested in the case of high magnetic fields and low temperatures. In this case, we have according to [100]

$$N_2(H) \propto \exp\left(\frac{be^2 H}{10R(T,0)\gamma\varepsilon k_B T}\right). \tag{10}$$

In accordance with the results of Refs. [99−101], the electrical conductivity $\sigma(H)$ depending on the applied magnetic field is proportional to the number of excited droplets $N_2$. However, an additional contribution to $\sigma(H)$ comes from the dependence of the relaxation time on magnetic field

$$\sigma(H) \propto N_2(H)\omega_0 \exp\left[-\frac{U_0(H)}{k_B T}\right], \tag{11}$$

where $\omega_0$ is the characteristic frequency of the electron motion within the ferron and $U_0(H)$ is the effective activation barrier, which an electron overcomes in the process of hopping between ferrons. This barrier evidently decreases with magnetic field due to (i) the growth of ferron sizes that gives rise to a decrease in the distance between ferrons and (ii) due to the aligning magnetic moments of these droplets with the increase of the magnetic field. As a result, we have

$$\sigma(H) \propto \exp\left[\frac{H}{k_B T}\left(\frac{b\gamma e^2}{10R(T,0)\gamma\varepsilon} + \left|\frac{\partial U_0}{\partial H}\right|\right)\right]. \tag{12}$$

Thus, we can write

$$\mu_{\text{eff}} = \frac{b\gamma e^2}{10R(T,0)\gamma\varepsilon} + \left|\frac{\partial U_0}{\partial H}\right|. \tag{13}$$

The experimental results of magnetoresistance are described well if one neglects $|U_0'(H)|$. In this case, $\mu_{\text{eff}} = \alpha(1 + T/T^*)^{-4/5}$ (see the solid (orange) curve in Fig. 3b), where $\alpha = gSe^2/10d\gamma\varepsilon(JzS^2)^{\frac{4}{5}}(\pi t)^{1/5}$. We take for estimates $t = 5$ meV, $g = 1$, $S = 1/2$, $J = 1.4$ K, $d \approx 4$ Å, $z = 12$, $\varepsilon = 6$ [102], and $\gamma = 2$. As a result, we get $T^* \approx 6$ K, $R(T, 0) \approx 1.6$–$1.7$ $d = 6.5$–$7$ Å when 2 K $< T <$ 6 K, and $\alpha \approx 15$. However, the best fit with the experimental data corresponds to $\alpha = 10$, which is a rather good agreement for the used simplified model.

The expression for $\sigma(H)$ in form (12) can be treated as a generalized Drude formula with $\sigma \propto \tau$, where $\tau = \tau(H)$ is an effective relaxation time for the ferron-assisted transport, which exhibits an exponential magnetic field dependence. In (12), the expression in parentheses may be denoted as $\mu_{\text{eff}}$, so that $\mu_{\text{eff}}H$ is the characteristic energy scale related to the applied magnetic field. In the framework of such a Drude-like approach, the contribution to the thermal conductivity related to ferrons can be described by Eq. (4). Note that ferron-assisted heat transfer involves the same scattering mechanisms as $\sigma(H)$. Therefore, $\kappa(H) \propto \tau(H)$ should also exhibit a similar behavior. The experimental data for $\kappa(H)$ indeed demonstrate the exponential magnetic field dependence, but with a smaller $\mu_{\text{eff}}$. It should be emphasized, that the experimental $C(H)$ curve exhibits an exponential *decrease* with the growth of $H$. According to Eqs. (4–6), the values of $\mu_{\text{eff}}$ for specific heat $C(H)$, thermal conductivity $\kappa(H)$, resistivity $\rho(H)$ and relaxation time $\tau(H)$ should obey the relation $\mu_{\text{eff}(\rho)} \approx \mu_{\text{eff}(\tau)} = \mu_{\text{eff}(C)} + \mu_{\text{eff}(\kappa)}$. Of course, the above reasoning is oversimplified and can be used only for qualitative estimates. Nevertheless, the experimental results $\mu_{\text{eff}(\rho)} \approx 1.8$ $\mu_B$ and $\mu_{\text{eff}(\tau)} \approx 2$ $\mu_B$ shown in Fig. 3b for the temperature $T = 1.9$ K, are in fairly good agreement.

At the same time, the values of $\mu_{\text{eff}(\rho)}$, $\mu_{\text{eff}(C)}$, and $\mu_{\text{eff}(\kappa)}$ estimated from experimental results by Eqs. (1)–(3), differ significantly (see Fig. 3b), and can be described by the inequality $\mu_{\text{eff}(C)} < \mu_{\text{eff}(\kappa)} < \mu_{\text{eff}(\rho)}$. We suppose that the distinction between $\mu_{\text{eff}(\rho)}$ and $\mu_{\text{eff}(\kappa)}$ comes, among other reasons, from the difference of the orientation of external magnetic field in these two experiments. Indeed, the charge transport $\rho(H)$ measurements were carried out using transverse magnetoresistance configuration ($\bm{H} \perp \bm{I}$), and thermal conductivity $\kappa(H)$ has been studied here in the longitudinal orientation of magnetic field ($\bm{H} \| \nabla T$). Taking into account, that the barrier height in the double-well potential decreases along the external magnetic field, we may assume that the activation process becomes easier in the field direction, leading to smaller $\mu_{\text{eff}(\kappa)}$ values (Fig. 3b). Note also that in comparison with $\rho_{\text{ferr}}(T)$ smaller values of $\rho_{\text{ferr}(\kappa)}(T)$ obtained here using the Wiedemann−Franz relation $1/\rho_{i(\kappa)} = \kappa_i/L_0 T$ (Fig. 3a) argue in favor of this explanation. However, we assume that the principal difference between the deduced effective moments $\mu_{\text{eff}(\rho)}$, $\mu_{\text{eff}(C)}$ and $\mu_{\text{eff}(\kappa)}$ may be related to the electron phase separation (Fig. 7), which leads to significant heterogeneity of the studied CeB$_6$ single crystals. Herewith, in the case of heat flow experiments both the dynamic charge stripes and ferrons in the bulk of CeB$_6$ crystals need to be considered only as inhomogeneities, which act as very effective scatterers, whereas the electrons in these intrinsic inclusions contribute also to the field-induced charge transport. In our view, the situation is strictly different in the case of the exponential behavior of heat capacity, which is likely very sensitive to field−induced ordering of these filamentary (stripes) and many−body (ferrons) inclusions in the unusual crystals. A similar scenario proposed recently for LuB$_{12}$, where the transition to a more ordered configuration of these structural defects have been observed at low temperatures and in high magnetic fields [103]. As it were, according to Eq. (4) the exponential behavior of the relaxation time is no longer determined by the ordering of stripes and ferrons in magnetic field, being the only characteristic of charge carrier scattering.

## 5. Conclusions

Studies of resistivity, thermal conductivity and specific heat at low temperature 1.8−7 K and in magnetic field up to 90 kOe allowed us to observe for the first time exponential field dependences $\rho(H)$, $\kappa^{-1}(H)$, $C(H) \sim \exp(-\mu_{\text{eff}}H/k_BT)$ of the charge transport and thermal characteristics in the so−called antiferroquadrupole phase of the archetypal heavy fermion $CeB_6$ hexaboride. It has been shown by magnetoresistance measurements that in the AFQ state the effective magnetic moment varies in the range $\mu_{\text{eff}}(T) = 1.4-1.9$ $\mu_B$, which is in fairly good accordance with $\mu_{\text{eff}(\tau)}(T) \approx 2$ $\mu_B$, obtained from the field dependence of relaxation time $\tau(H)$ in heat capacity and thermal conductivity experiments. We proposed a phenomenological model, which allowed us to attribute the magnetic moments to spin droplets (ferrons), which develop in the AFQ phase of $CeB_6$. Moreover, an electron phase separation on the nanoscale in the form of dynamic charge stripes was discovered at $T = 30$ K from the analysis of X-ray diffraction experiments using the maximum entropy method. We argue that the Jahn−Teller collective mode of $B_6$ clusters, the ferrodistortive effect, is responsible for the formation of dynamic charge stripes, which based on the $2p$ states of boron. The JT mode induces transverse quasi-local vibrations of Ce ions' pairs and triples that provide $4f-5d$ spin fluctuations and produce spin-polarized $5d$ states (ferrons) in the matrix of $CeB_6$.

## Acknowledgements

The work in Prokhorov General Physics Institute of RAS was supported by the Russian Science Foundation, Project No. 23−22−00297, and partly performed using the equipment of the Shared Research Centers of FSRC 'Crystallography and Photonics' of NRC KI and Lebedev Physical Institute of RAS. The authors are grateful to A.V. Semeno and V.V. Glushkov for helpful discussions. S.G. and K.F. acknowledge the support of the Slovak Research and Development Agency under contract No. APVV−23−0226.

A.N. Azarevich[1], O.N. Khrykina[1,2], N.B. Bolotina[2], V.G. Gridchina[1,2], A.V. Bogach[1], S. V. Demishev[3, 1], V.N. Krasnorussky[3, 1], S. Yu. Gavrilkin[4], A. Yu. Tsvetkov[4], N.Yu. Shitsevalova[5], V.V. Voronov[1], K.I. Kugel[6], A.L. Rakhmanov[6], S. Gabani[7], K. Flachbart[7], N.E. Sluchanko[1]

[1]*Prokhorov General Physics Institute of the Russian Academy of Sciences, Vavilova 38, 119991, Moscow, Russia*

[2]*National Research Center Kurchatov Institute, Moscow, 123182 Russia*

[3]*Vereshchagin Institute for High Pressure Physics of the Russian Academy of Sciences, 142190, Troitsk, Moscow, Russia*

[4]*Lebedev Physical Institute, Russian Academy of Sciences, Moscow, Leninsky 53, 119991 Russia*

[5]*Frantsevich Institute for Problems of Materials Science, National Academy of Sciences of Ukraine, 03680 Kyiv, Ukraine*

[6]*Institute for Theoretical and Applied Electrodynamics of Russian Academy of Sciences, Izhorskaya str. 13, 125412 Moscow, Russia*

[7]*Institute of Experimental Physics of the Slovak Academy of Sciences, Watsonova 47, SK-04001 Košice, Slovakia*

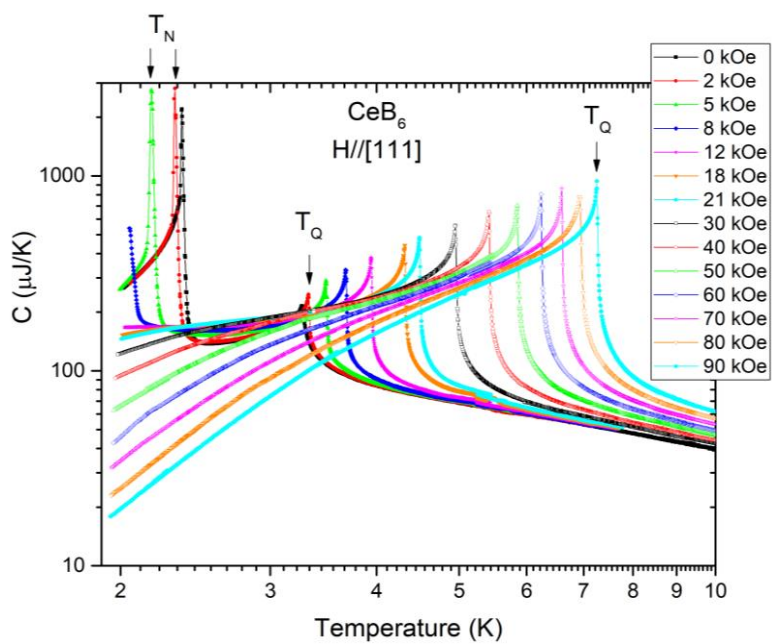

**Fig. S1.** Temperature dependences of heat capacity in external magnetic field *H*//[111] up to 90 kOe. $T_N$ and $T_Q$ denote the temperatures of the magnetic phase transitions.

**Table S1.** Crystallographic characteristics, details of X-ray diffraction experiments with the $CeB_6$ single crystals.

| Sample | $CeB_6$ | |
|---|---|---|
| $T$, K | 30 | 200 |
| Space group, $Z$ | $Pm\text{–}3m$, 1 | |
| Lattice parameter $a_{cub}$, Å | 4.1342(1) | 4.1385(1) |
| Radiation type; $\lambda$, Å | $AgK_\alpha$, 0.56087 | |
| Maximum linear size of the sample, mm | 0.204 | |
| Diffractometer | XtaLAB Synergy-DW HyPix Arc 150° | |
| Scan mode | $\Omega$ | |
| Absorption correction | polyhedron | |
| Absorption coefficient $\mu$, mm$^{-1}$; $T_{min}$, $T_{max}$ | 8.332; 0.686, 0.858 | 8.336; 0.412, 0.794 |
| $\theta_{max}$, deg | 45.28 | 74.38 |
| Limits of $h$; $k$; $l$ | $-10 \le h \le 10$; $-10 \le k \le 10$; $-9 \le l \le 10$ | $-13 \le h \le 14$; $-14 \le k \le 14$; $-14 \le l \le 14$ |
| Number of reflections: observed; with $I > 3\sigma I$; independent; $R_{int}$, % | 22281; 16533; 163; 7.48 | 20081; 10696; 200; 5.83 |
| Refinement method | Least squares on $F$ | |
| Numbers of refined parameters | 6 | |
| Extinction type | Type 1, Lorentzian | |
| $R(|F|) / wR(|F|)$, % | 0.86 / 1.00 | 1.46 / 2.01 |
| Goodness of fit, $S$ | 1.03 | 1.03 |
| $\Delta\rho_{min}/\Delta\rho_{max}$, e/Å$^3$ | –0.90 / 0.90 | –1.12 / 2.37 |
| Programs | CrysAlis Pro v.43.92, Jana2006 | |

**Table S2.** Small JT-induced static distortions of the CeB$_6$ crystal lattice.

| Chemical formula | CeB$_6$ | |
|---|---|---|
| $T$, K | 30 | 200 |
| $a$, Å | 4.1333(1) | 4.1385(1) |
| $b$, Å | 4.1360(1) | 4.1395(1) |
| $c$, Å | 4.1333(1) | 4.1375(1) |
| $\alpha$, deg | 89.991(2) | 90.004(2) |
| $\beta$, deg | 90.000(2) | 90.000(2) |
| $\gamma$, deg | 89.965(2) | 89.989(2) |

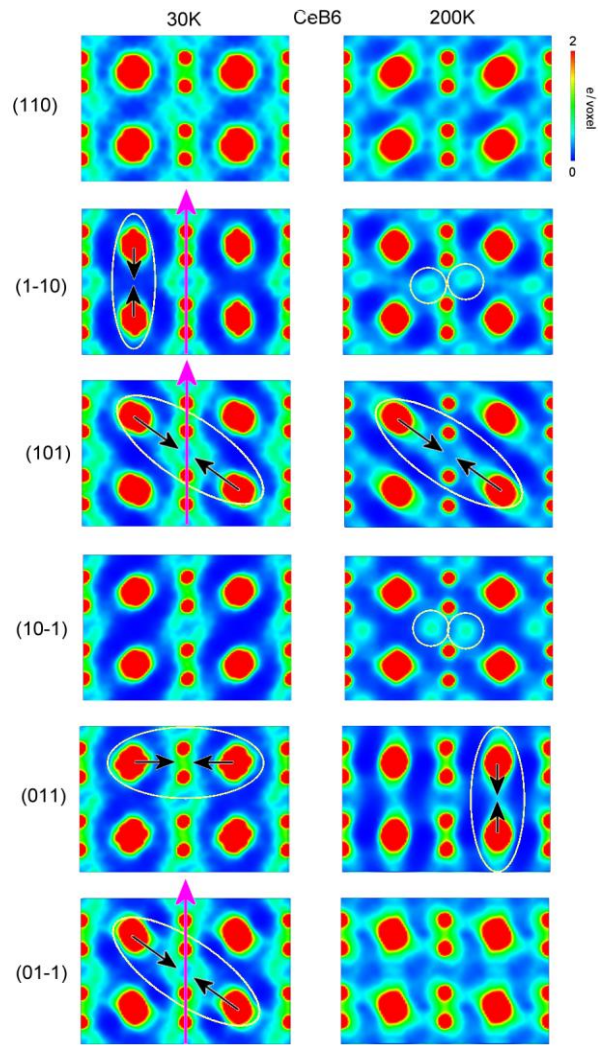

**Fig. S2.** MEM maps of CeB$_6$ in the {110} planes at the temperatures of 30 K (left column) and 200 K (right column). Electron density (ED) peaks are cut off at a height of 2 e/voxel to highlight fine details of the ED distribution in the lattice interstices. Large red circles – Ce positions; small red circles – positions B. Shades of green highlight the interstitial electron density in the framework of boron atoms. Yellow rings highlight the sub-structural charge density waves (s-CDW). Yellow ellipses with short black arrows inside show a vibrationally coupled Ce-Ce pairs, and long pink arrows mark the dynamic charge stripes.